\documentclass[usenatbib]{mn2e}

\usepackage{amsmath}
\usepackage{epsf}

\newcommand\bfT{\mathbf{T}}

\newcommand\bB{{\bmath B}}
\newcommand\be{{\bmath e}}
\newcommand\bE{{\bmath E}}

\newcommand\bJ{{\bmath J}}

\newcommand\bu{{\bmath u}}
\newcommand\bv{{\bmath v}}
\newcommand\bL{{\bmath L}}

\newcommand\bX{{\bmath X}}
\newcommand\bY{{\bmath Y}}

\newcommand\bnabla{{\bmath\nabla}}

\newcommand\rma{\mathrm{a}}
\newcommand\rmb{\mathrm{b}}
\newcommand\rmd{\mathrm{d}}

\newcommand\rms{\mathrm{s}}

\newcommand\rmw{\mathrm{w}}
\newcommand\cs{c_\mathrm{s}}
\newcommand\zs{z_\mathrm{s}}
\newcommand\f{\frac}
\newcommand\p{\partial}

\title[Jet launching from accretion discs]
{Jet launching from accretion discs in the local approximation}

\author[Gordon I. Ogilvie]{Gordon I. Ogilvie\\
Department of Applied Mathematics and Theoretical Physics,
University of Cambridge, Centre for Mathematical Sciences,\\
Wilberforce Road, Cambridge CB3 0WA}

\begin{document}

\maketitle

\label{firstpage}
 
\begin{abstract}
  The acceleration of an outflow along inclined magnetic field lines
  emanating from an accretion disc can be studied in the local
  approximation, as employed in the computational model known as the
  shearing box.  By including the slow magnetosonic point within the
  computational domain, the rate of mass loss in the outflow can be
  calculated.  The accretion rates of mass and magnetic flux can also
  be determined, although some effects of cylindrical geometry are
  omitted.  We formulate a simple model for the study of this problem
  and present the results of one-dimensional numerical simulations and
  supporting calculations.  Quasi-steady solutions are obtained for
  relatively strong poloidal magnetic fields for which the
  magnetorotational instability is suppressed.  In this regime the
  rate of mass loss decreases extremely rapidly with increasing field
  strength, or with decreasing surface density or temperature.  If the
  poloidal magnetic field in an accretion disc can locally achieve an
  appropriate strength and inclination then a rapid burst of ejection
  may occur.  For weaker fields it may be possible to study the
  launching process in parallel with the magnetorotational
  instability, but this will require three-dimensional simulations.
\end{abstract}


\section{Introduction}

In a well known paper, \citet{1982MNRAS.199..883B} presented a
mechanism by which jets or winds can be launched from accretion discs.
Consider a disc that is threaded by a poloidal magnetic field and is
sufficiently ionized for ideal magnetohydrodynamics (MHD) to be a good
approximation.  Above the surface of the disc, where the magnetic
field is dynamically dominant, gas tends to rotate with the same
angular velocity as the part of the disc to which is magnetically
connected; its motion in the meridional plane also tends to be
parallel to the magnetic field.  Above a thin Keplerian disc, the net
acceleration of gas parallel to the field is directed away from the
disc if the poloidal field is inclined at more than $30^\circ$ to the
vertical, when both the centrifugal and gravitational forces are taken
into account.

\begin{figure}
\centerline{\epsfysize8cm\epsfbox{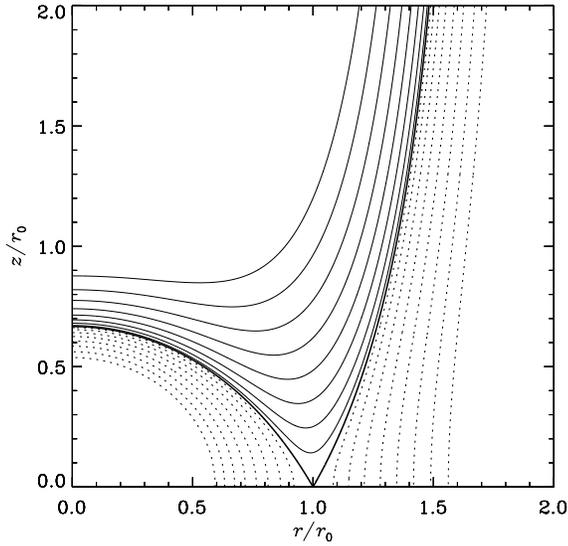}}
\caption{Contours of the effective potential experienced by matter
that is forced to rotate with the Keplerian angular velocity at radius
$r_0$.  The contour values are unequally spaced.  Dotted contours
correspond to values lower than that of the saddle point.}
\label{f:bp}
\end{figure}

Fig.~\ref{f:bp}, based on Fig.~1 of \citet{1982MNRAS.199..883B},
illustrates the effective potential experienced by gas that is forced
to corotate with the angular velocity of a Keplerian orbit of radius
$r_0$ in the plane $z=0$.  Near the saddle point of the effective
potential at this same location, acceleration away from the disc can
occur if the inclination exceeds $30^\circ$ either to the left or to
the right.  Indeed, the picture is symmetrical near the saddle point.
Globally, however, only flows that are directed away from the rotation
axis are suitable for launching jets.

\section{Local approximation}

\subsection{Introduction}

\begin{figure}
\centerline{\epsfysize4.5cm\epsfbox{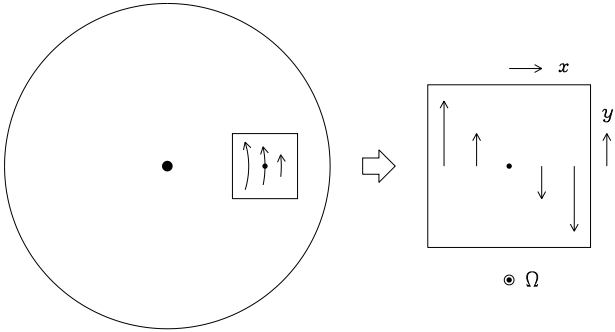}}
\medskip
\caption{The shearing sheet, or local approximation.  Coordinates $x$,
$y$ and $z$ are measured in the radial, azimuthal and vertical
directions.}
\label{f:sheet}
\end{figure}

The shearing sheet \citep[Fig.~\ref{f:sheet};][]{1965MNRAS.130..125G}
is a widely adopted local model of an astrophysical disc.  The sheet
is centred on a reference point that follows a circular orbit around
the central mass, and the frame rotates with the angular velocity
$\Omega$ of this orbit.  The differential rotation of the disc is
represented locally as a uniform parallel shear flow in the rotating
frame.  The local approximation has two symmetries not present in the
global system: it is invariant under translation in the $x$ direction
(allowing for a Galilean transformation in the $y$ direction to
correct for the velocity shift) and under rotation by $180^\circ$
about the $z$ axis.  The latter symmetry might appear to make the
shearing sheet unsuitable for studying jet launching, because it fails
to distinguish between radially inward and outward directions;
however, as noted above, this symmetry is indeed present in the
problem of \citet{1982MNRAS.199..883B} close to the disc.

For simplicity, we assume in this paper that the disc orbits in a
Newtonian potential, giving rise to Keplerian rotation in the absence
of other radial forces, and that the gas is isothermal, its pressure
and density being related by $p=\cs^2\rho$, with a uniform isothermal
sound speed $\cs$.  The gas satisfies the equation of motion,
\begin{eqnarray}
  &&\rho\left(\f{\p\bu}{\p t}+\bu\cdot\bnabla\bu+2\Omega\,\be_z\times\bu\right)=-\rho\bnabla\Phi-\bnabla p\nonumber\\
  &&\qquad\qquad+\bJ\times\bB+\bnabla\cdot\bfT,
\end{eqnarray}
the equation of mass conservation,
\begin{equation}
  \f{\p\rho}{\p t}+\bnabla\cdot(\rho\bu)=0,
\end{equation}
the induction equation,
\begin{equation}
  \f{\p\bB}{\p t}=\bnabla\times(\bu\times\bB-\eta\bnabla\times\bB),
\end{equation}
and the solenoidal condition
\begin{equation}
  \bnabla\cdot\bB=0.
\end{equation}
Here $\Phi=\tfrac{1}{2}\Omega^2(z^2-3x^2)$ is the effective (tidal)
potential in the local approximation, which comes from expanding the
sum of the gravitational and centrifugal potentials to second order in
the distance from the centre of the sheet.  Also
$\bJ=\mu_0^{-1}\bnabla\times\bB$ is the electric current density and
$T_{ij}=\rho\nu(u_{i,j}+u_{j,i})+\rho(\nu_\rmb-\tfrac{2}{3}\nu)u_{k,k}\delta_{ij}$
is the viscous stress tensor.  We assume that the kinematic shear and
bulk viscosities $\nu$ and $\nu_\rmb$ and the magnetic diffusivity
$\eta$ are uniform.

The hyperbolic contours of $\Phi$ in the $xz$ plane agree with those
of the effective potential plotted in Fig.~\ref{f:bp} close to the
saddle point and reproduce the critical inclination of $30^\circ$.
The reason for this is of course that $\Phi$ is the effective
potential experienced by matter that is forced to rotate with the
Keplerian angular velocity $\Omega$ of the reference point.

The basic state of the shearing sheet in the absence of magnetic
fields consists of the Keplerian shear flow $\bu=-\tfrac{3}{2}\Omega
x\,\be_y$ together with the hydrostatic density distribution
$\rho=\rho_0\exp(-z^2/2H^2)$, where $\rho_0$ is a constant and
$H=\cs/\Omega$ is the isothermal scaleheight.  The departure from the
basic Keplerian flow is denoted by $\bv=\bu+\tfrac{3}{2}\Omega
x\,\be_y$.

This system of equations can be solved numerically in a finite
shearing box \citep[e.g.][]{1995ApJ...440..742H}, in which case
shearing-periodic horizontal boundary conditions apply to the solution
(to $\bv$ rather than $\bu$), while various vertical boundary
conditions are permissible.  The shearing box has been widely employed
in treatments of the magnetorotational instability (MRI) \citep[][and
references therein]{1998RvMP...70....1B}, but has not been used for
studies of jet launching.  [The recent work of
\citet{2009ApJ...691L..49S} and \citet{2010ApJ...718.1289S} finds mass
loss from the computational domain but does not consider the
systematically inclined fields relevant for the mechanism of
\citet{1982MNRAS.199..883B}.]

\subsection{Vertical boundary conditions}

\begin{figure}
\centerline{\epsfysize7cm\epsfbox{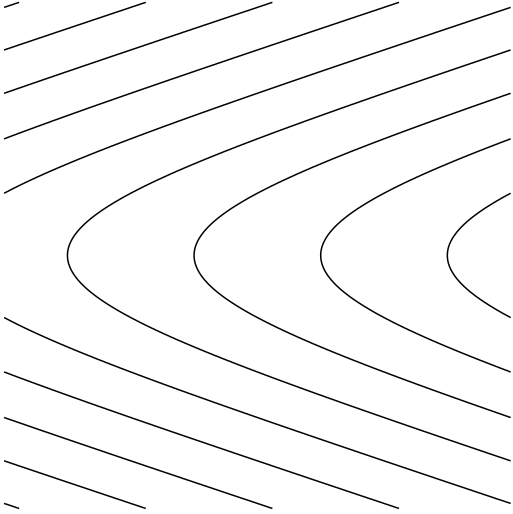}}
\bigskip
\caption{Geometry of the poloidal magnetic field in the $(x,z)$ plane
  that is favourable for jet launching in the local approximation.
  The field lines could equally well bend the other way.}
\label{f:bend}
\end{figure}

For any choice of vertical boundary conditions at $z=\pm Z$, and with
shearing-periodic horizontal boundary conditions, the horizontal
average over the box of $B_z$ is independent of $z$ and $t$, and is
determined by the initial conditions.  Within the local approximation,
the type of poloidal magnetic field configuration that is favourable
for jet launching consists of a uniform vertical field $B_z$ together
with a radial field $B_x$ that is odd in $z$ and tends to a non-zero
constant at large $z$; this represents a field that is straight,
inclined, current-free and therefore force-free in the low-density gas
at large $|z|$ and bends symmetrically as it passes through the disc,
producing a radial Lorentz force through the azimuthal current $J_y$
(Fig.~\ref{f:bend}).

In this type of study, where a thin disc is to be connected to a jet,
the question arises of which quantities are determined by the disc and
which by the jet \citep[see][and references
therein]{2001ApJ...553..158O}.  Efficient outflows pass through a slow
magnetosonic point not far above the surface of the disc and through
an Alfv\'en point much higher up \citep{1996epbs.conf..249S}.  It is
convenient to think of matching the disc to the jet a small distance
above the slow magnetosonic point, in a region where the magnetic
field is predominantly poloidal and approximately force-free.  Local
studies of the disc can focus on the dynamics of the disc and the
passage of the outflow through the slow point, which determines the
rate of mass loss.  However, the rate of angular momentum loss, and
therefore the magnetic torque on the disc, are determined by the
passage of the outflow through the Alfv\'en point, which usually lies
outside the domain of the local approximation.  Since the magnetic
torque is proportional to the azimuthal magnetic field, the value of
$B_y$ in the matching region is determined by the jet region, not the
disc region.  Given the distribution of poloidal magnetic flux on the
midplane of the disc, the global shape of the poloidal magnetic field
is also determined by the jet region.  It is therefore appropriate to
impose on the local disc the value of $B_z$ and the values of $B_x$
and $B_y$ in the matching region.  This can be done by a novel type of
boundary condition in which the horizontal components of the magnetic
field are specified at the vertical boundaries, with equal and
opposite values at the top and bottom because of the desired symmetry.
At the same time, the vertical boundary conditions should leave the
density and the velocity free to evolve according to the dynamics of
the outflow that develops.

\subsection{Evolution of the horizontal momentum}

Let angle brackets denote a horizontal average over the box.  Then, by
rewriting the horizontal components of the equation of motion in
conservative form and integrating over the box, we obtain
\begin{eqnarray}
  \lefteqn{\p_t\int_{-Z}^Z\langle\rho v_x\rangle\,\rmd z-2\Omega\int_{-Z}^Z\langle\rho v_y\rangle\,\rmd z}&\nonumber\\
  &&=\left[\bigg\langle\f{B_xB_z}{\mu_0}-\rho v_xv_z+T_{xz}\bigg\rangle\right]_{-Z}^Z,
\end{eqnarray}
\begin{eqnarray}
  \lefteqn{\p_t\int_{-Z}^Z\langle\rho v_y\rangle\,\rmd z+\f{1}{2}\Omega\int_{-Z}^Z\langle\rho v_x\rangle\,\rmd z}&\nonumber\\
  &&=\left[\bigg\langle\f{B_yB_z}{\mu_0}-\rho v_yv_z+T_{yz}\bigg\rangle\right]_{-Z}^Z.
\label{ame}
\end{eqnarray}
In most shearing-box simulations the right-hand sides of these
equations are negligible or zero, with the result that the mean
horizontal momentum of the box executes an epicyclic oscillation of
constant amplitude, which can be made to vanish by a suitable choice
of initial conditions.  Here, however, we wish to impose non-vanishing
time-independent magnetic stresses at the vertical boundaries.  These
provide source terms for the epicyclic oscillation.  If the initial
conditions are chosen carefully, a non-oscillatory solution is
possible, in which the $xz$ stress is matched by a mean flow in the
$y$ direction and the $yz$ stress by a mean flow in the $x$ direction.
The first represents a departure from Keplerian rotation induced by
the radial Lorentz force of a poloidal magnetic field that bends
through the disc, while the second represents an accretion flow driven
by an imposed magnetic torque.  For most initial conditions, however,
a free epicyclic oscillation of constant amplitude and phase will be
superimposed on this non-oscillatory solution.  (In fact, depending on
the way that mass loss from a finite computational domain is treated,
the amplitude may \emph{not} be constant, as is discussed below.)

\section{Horizontally uniform solutions}

We do not attempt the numerical solution of this system in multiple
dimensions in this paper.  Indeed, simulations of the MRI including
both vertical gravity and a net vertical magnetic field present
serious numerical and interpretative difficulties
(\citealt{2000ApJ...534..398M}; S.~Fromang, private communication;
G.~Lesur, private communication; J.~M.~Stone, private
communication).  For the remainder of this paper we focus on
horizontally uniform solutions that depend only on $z$ and $t$.  These
are admissible because of the horizontal translational symmetry of the
local approximation. Then we have

\begin{equation}
  \rho\left(\f{\p v_x}{\p t}+v_z\f{\p v_x}{\p z}-2\Omega v_y\right)=J_yB_z+\f{\p}{\p z}\left(\rho\nu\f{\p v_x}{\p z}\right),
\end{equation}
\begin{equation}
  \rho\left(\f{\p v_y}{\p t}+v_z\f{\p v_y}{\p z}+\f{1}{2}\Omega v_x\right)=-J_xB_z+\f{\p}{\p z}\left(\rho\nu\f{\p v_y}{\p z}\right),
\end{equation}
\begin{eqnarray}
  &&\rho\left(\f{\p v_z}{\p t}+v_z\f{\p v_z}{\p z}\right)=-\rho\Omega^2z-\f{\p p}{\p z}\nonumber\\
  &&\qquad+J_xB_y-J_yB_x+\f{\p}{\p z}\left[\rho(\nu_\mathrm{b}+{\textstyle\f{4}{3}}\nu)\f{\p v_z}{\p z}\right],
\end{eqnarray}
\begin{equation}
  \f{\p\rho}{\p t}+\f{\p}{\p z}(\rho v_z)=0,
\end{equation}
\begin{equation}
  \f{\p B_x}{\p t}=\f{\p E_y}{\p z},
\end{equation}
\begin{equation}
  \f{\p B_y}{\p t}=-\f{3}{2}\Omega B_x-\f{\p E_x}{\p z},
\end{equation}
\begin{equation}
  \f{\p B_z}{\p t}=0,
\end{equation}
\begin{equation}
  \f{\p B_z}{\p z}=0,
\end{equation}
with
\begin{equation}
  E_x=v_zB_y-v_yB_z+\eta J_x,
\end{equation}
\begin{equation}
  E_y=v_xB_z-v_zB_x+\eta J_y,
\end{equation}
\begin{equation}
  \mu_0J_x=-\f{\p B_y}{\p z},
\end{equation}
\begin{equation}
  \mu_0J_y=\f{\p B_x}{\p z}.
\end{equation}
Clearly $B_z$ is a constant, which becomes a parameter of the problem.
The solution of these equations also contains a free epicyclic
oscillation of arbitrary amplitude and phase.  That is, the equations
are invariant under the transformation $v_x\mapsto
v_x+2a\sin[\Omega(t-\tau)]$, $v_y\mapsto v_y+a\cos[\Omega(t-\tau)]$,
for any real numbers $a$ and $\tau$.\footnote{The electric field $\bE$
  undergoes a non-trivial transformation.  In fact $\bE$ is not the
  total electric field because it does not include the contribution
  (proportional to $x$) from induction by the Keplerian shear flow.}
In general, the initial conditions will excite this motion at some
amplitude.

In the limit of ideal MHD ($\nu=\nu_\rmb=\eta=0$) the equations form a
hyperbolic system, with wave speeds $v$ given by
\begin{equation}
  [(v-v_z)^2-v_{\rma z}^2][(v-v_z)^4-(\cs^2+v_\rma^2)(v-v_z)^2+\cs^2v_{\rma z}^2]=0,
\label{wavespeeds}
\end{equation}
where $\bv_\rma=(\mu_0\rho)^{-1/2}\bB$ is the Alfv\'en velocity.
The six solutions are, of course, the Alfv\'en waves and the fast and slow
magnetoacoustic waves.

If the upper boundary $z=Z$ is located above the slow magnetosonic
point but below the Alfv\'en point, then two of the six waves are
directed into the computational domain and the remaining four waves
are directed outward.  Therefore two boundary conditions should be
applied to the six dependent variables $(v_x,v_y,v_z,\rho,B_x,B_y)$
there, and we do this by specifying the values of $B_x$ and $B_y$, as
discussed above.

\section{Wave speed limiter}

A difficulty arises in solving the above system of equations if the
density becomes very small above the disc.  The very large Alfv\'en
speed means that a very short timestep is required for stability
unless an implicit integration scheme is adopted.  It is possible to
limit the wave speeds in a way that, to some extent, mimics the
effects of Einsteinian relativity.  This
can be done by introducing a displacement current, replacing the
definitions of $J_x$ and $J_y$ by
\begin{equation}
  \mu_0J_x+\f{1}{c^2}\f{\p E_x}{\p t}=-\f{\p B_y}{\p z},
\end{equation}
\begin{equation}
  \mu_0J_y+\f{1}{c^2}\f{\p E_y}{\p t}=\f{\p B_x}{\p z},
\end{equation}
where $c$ mimics the speed of light, but is a free parameter of the
model.  (Optionally, $J_z$ and $E_z$ can be included as well.)  When
$\nu=\nu_\rmb=\eta=0$ we now have a hyperbolic system of equations.
It is neither Galilean-invariant nor Lorentz-invariant, but this is
convenient for our purposes because the added terms limit the speed of
waves relative to the coordinate system (and the numerical grid) while
having no effect on steady flows and their critical points.  We use
the wave speed limiter mainly as a convenient way of reaching steady
solutions using fewer timesteps; these are then also steady solutions
of the original problem.  [The more complicated method of
\citet{2000ApJ...534..398M} does not appear to have this property,
although it may have other desirable features.]  In practice the wave
speeds are limited to $\max(c,\cs+|u_z|)$.

Provided that $\eta$ is uniform, the system is still invariant under
the addition of an epicyclic motion, although the transformation law
of the electric field is modified.

\section{Quasi-steady solutions}

\subsection{ODE system}

The system does not in fact admit truly steady solutions with an
outflow unless a source of mass is included.  In this case we obtain
the system of ordinary differential equations (ODEs)
\begin{equation}
  \rho\left(v_z\f{\rmd v_x}{\rmd z}-2\Omega v_y\right)=\f{\rmd}{\rmd z}\left(\f{B_xB_z}{\mu_0}+\rho\nu\f{\rmd v_x}{\rmd z}\right),
\end{equation}
\begin{equation}
  \rho\left(v_z\f{\rmd v_y}{\rmd z}+\f{1}{2}\Omega v_x\right)=\f{\rmd}{\rmd z}\left(\f{B_yB_z}{\mu_0}+\rho\nu\f{\rmd v_y}{\rmd z}\right),
\end{equation}
\begin{eqnarray}
  &&\rho v_z\f{\rmd v_z}{\rmd z}=-\rho\Omega^2z\nonumber\\
  &&+\f{\rmd}{\rmd z}\left[-p-\f{(B_x^2+B_y^2)}{2\mu_0}+\rho(\nu_\mathrm{b}+{\textstyle\f{4}{3}}\nu)\f{\rmd v_z}{\rmd z}\right],
\end{eqnarray}
\begin{equation}
  \f{\rmd}{\rmd z}(\rho v_z)=\dot\rho,
\end{equation}
\begin{equation}
  0=\f{\rmd}{\rmd z}\left(v_xB_z-v_zB_x+\eta\f{\rmd B_x}{\rmd z}\right),
\end{equation}
\begin{equation}
  0=-\f{3}{2}\Omega B_x+\f{\rmd}{\rmd z}\left(v_yB_z-v_zB_y+\eta\f{\rmd B_y}{\rmd z}\right),
\end{equation}
\begin{equation}
  \f{\rmd B_z}{\rmd z}=0,
\end{equation}
where $\dot\rho(z)$ is the mass source.

One possibility is to provide a mass source only on the midplane
$z=0$, so that $\dot\rho\propto\delta(z)$.  The ODEs can then be
solved in $z>0$, with $\dot\rho=0$.  This procedure resembles the way
in which steady accretion discs are treated without accounting for the
(slow) increase in the mass of the central object.  Alternatively,
mass can be replenished at a rate proportional to the local density,
so that $\dot\rho=\gamma\Omega\rho$, where $\gamma$ is a small
positive constant whose value can be adjusted to obtain a steady
solution.  In either case, mass is added to the solution at a rate
that balances the outflow.

The solution is expected to be symmetrical about the midplane, with
$\rho$, $v_x$ and $v_y$ being even in $z$ while $v_z$, $B_x$ and $B_y$
are odd.  In the case of a mass source localized on the midplane,
however, $v_z$ is non-zero (and discontinuous) at $z=0$.

The ideal MHD problem (in which $\nu=\nu_\mathrm{b}=\eta=0$) has
critical points wherever the wave speeds $v$ given by
equation~(\ref{wavespeeds}) vanish.  We assume that $\rho>0$ and
$v_z>0$ for all $z>0$.  We expect that the flow will accelerate beyond
the slow magnetosonic speed a few scale heights above the midplane, if
the inclination of the poloidal field is appropriate.  In a successful
outflow the Alfv\'en and fast magnetosonic speeds will also be
exceeded, but this will usually happen at a greater distance beyond
the range of validity of the local approximation.  We therefore
consider only the slow magnetosonic critical point here.

We apply the following boundary conditions.  At the top of the domain,
$z=Z$, representing the `surface' of the disc, we prescribe the values
of $B_x$ and $B_y$ to be $B_{x\rms}$ and $B_{y\rms}$.  The constant
$B_z$ is also given as an input parameter.  The inclination angle $i$
of the poloidal magnetic field is defined through $B_{x\rms}=B_z\tan
i$.  At the midplane, $z=0$, the symmetry conditions $B_x=B_y=0$ apply,
as well as $v_z=0$ in the case of the distributed mass source.
Finally, we require the surface density
\begin{equation}
  2\int_0^Z\rho\,\rmd z=\Sigma
\label{sigma}
\end{equation}
to equal a given value, which is one of the parameters of the model.

Of particular interest as an output of the calculation is the rate of
mass loss per unit area from the upper surface of the disc,
\begin{equation}
  \dot m_\rmw=\rho v_z\Big|_{z=Z}.
\end{equation}
The timescale for depleting the mass of the disc is $\Sigma/(2\dot
m_\rmw)$.

Radial mass transport can also be induced in this model, despite the
symmetries of the local approximation.  Although the $xy$ components
of the viscous, Reynolds and Maxwell stresses are independent of $x$
and therefore do not drive an accretion flow, the magnetic stress
$B_yB_z/\mu_0$ acting at the vertical boundaries can do so.  In a
steady state, the integrated azimuthal momentum equation~(\ref{ame})
relates the radial transport velocity of mass, $v_m$, defined through
\begin{equation}
  \Sigma v_m=\int_{-Z}^Z\rho v_x\,\rmd z,
\end{equation}
to the $yz$ stresses acting at $z=\pm Z$.

Furthermore, the radial transport velocity of poloidal magnetic flux,
$v_\psi$, is given by
\begin{equation}
  E_y=v_xB_z-v_zB_x+\eta\f{\p B_x}{\p z}=v_\psi B_z.
\end{equation}
The physical significance of this quantity is discussed by
\citet{2001ApJ...553..158O} and Guilet \& Ogilvie (MNRAS, submitted).
It contains both an advective contribution, from motion in the
meridional plane across the field lines, and a diffusive contribution.

\citet{2001ApJ...553..158O} solved a closely related problem for the
vertical structure of magnetized discs with outflows.  There are
several differences.  In that work, the problem was separated into an
optically thick disc and an optically thin atmosphere.  In the disc,
the vertical velocity was neglected and the temperature was determined
according to an energy equation including viscous and resistive
heating and radiative diffusion.  In the atmosphere, an isothermal
outflow was computed along rigid magnetic field lines.  Another
difference is that the vertical transport of momentum by viscosity was
neglected, while the radial transport of angular momentum was modelled
in a way that goes beyond the standard local approximation.

\subsection{Ideal MHD problem}
\label{s:ideal}

We consider here the ideal MHD problem ($\nu=\nu_\rmb=\eta=0$) with
mass replenishment at the midplane.  Assuming that the inclination of
the poloidal magnetic field exceeds $30^\circ$, we expect to find a
slow magnetosonic point at some height $z=\zs$.

Given that $\dot m_\rmw=\rho v_z$ and $E_y=v_xB_z-v_zB_x$ are both
independent of $z$, we can choose our vector of dependent variables to
be
\begin{equation}
  \bX=\begin{bmatrix}v_y\\v_z\\B_x\\B_y\end{bmatrix}.
\end{equation}
The quantities $\rho$ and $v_x$ follow from the values of $\dot
m_\rmw$ and $E_y$, which have the nature of eigenvalues, while $B_z$
is a constant parameter.  The remaining ODEs can be cast in the form
\begin{equation}
  \mathbf{A}\f{\rmd\bX}{\rmd z}=\bY,
\end{equation}
where
\begin{equation}
  \mathbf{A}=\begin{bmatrix}0&B_xv_z&v_z^2-v_{\rma z}^2&0\\v_z&0&0&-\f{B_z}{\mu_0\rho}\\0&v_z-\f{\cs^2}{v_z}&\f{B_x}{\mu_0\rho}&\f{B_y}{\mu_0\rho}\\-B_z&B_y&0&v_z\end{bmatrix}
\end{equation}
and
\begin{equation}
  \bY=\begin{bmatrix}2\Omega B_zv_y\\-\f{1}{2}\Omega v_x\\-\Omega^2z\\-\f{3}{2}\Omega B_x\end{bmatrix}.
\end{equation}
At the slow magnetosonic point where
\begin{equation}
  v_z=\left\{\f{1}{2}(\cs^2+v_\rma^2)-\left[\f{1}{4}(\cs^2+v_\rma^2)^2-\cs^2v_{\rma z}^2\right]^{1/2}\right\}^{1/2},
\end{equation}
the matrix $\mathbf{A}$ is singular.  For a regular solution that
passes smoothly through the slow point with finite derivatives, we
require $\bL\bY=0$, where
\begin{equation}
  \bL=\begin{bmatrix}B_x&B_yB_z&B_z^2-\mu_0\rho v_z^2&B_yv_z\end{bmatrix}
\end{equation}
is the null left eigenvector of $\mathbf{A}$ satisfying
$\bL\mathbf{A}=\mathbf{0}$ at this point.  The regularity condition is
therefore
\begin{equation}
\label{regularity}
  B_yB_zv_x-4B_xB_zv_y+3B_xB_yv_z+2\Omega z(B_z^2-\mu_0\rho v_z^2)=0.
\end{equation}

A method of solution is as follows.  The quantities $z_\rms$ and $E_y$
are guessed, as well as the values of $B_x$, $B_y$ and $\rho$ at the
slow point.  The value of $v_z$ at the slow point follows by
definition and hence the quantity $\dot m_\rmw$.  The regularity
condition determines the value of $v_y$.  This is sufficient
information to integrate the ODEs in each direction away from the slow
point.  Away from the slow point and other critical points, the matrix
$\mathbf{A}$ can be inverted to find $\rmd\bX/\rmd z$ from $\bY$.  We
integrate down to $z=0$ and require that $B_x=B_y=0$ there.  We also
integrate up to $z=Z$ and require $B_x$ and $B_y$ to equal specified
values there.  A fifth condition is given by the mass
integral~(\ref{sigma}).  Using Newton--Raphson iteration, the five
guessed quantities are adjusted to meet the five conditions.

This method fails if the Alfv\'en point is encountered below $z=Z$,
because then the matrix $\mathbf{A}$ is singular and the solution
cannot be made to pass smoothly through the critical point and match
the specified values of $B_x$ and $B_y$ at $z=Z$.

A reduced version of the method is to assume that $B_x$ and $B_y$ have
their limiting values already at the slow point.  In that case no
upward integration is required.  There are only three quantities to
guess and three conditions to be met.

\subsection{Simplified version}
\label{s:simplified}

A simplified version of the problem is obtained by neglecting $v_x$,
$v_z$ and $B_y$ and solving the equations
\begin{equation}
  -2\Omega v_y=\f{B_z}{\mu_0\rho}\f{\rmd B_x}{\rmd z},
\end{equation}
\begin{equation}
  0=-\Omega^2z-\f{1}{\rho}\f{\rmd}{\rmd z}\left(p+\f{B_x^2}{2\mu_0}\right),
\end{equation}
\begin{equation}
  0=-\f{3}{2}\Omega B_x+B_z\f{\rmd v_y}{\rmd z}.
\end{equation}
This procedure is equivalent to assuming an isorotational
configuration with a purely poloidal magnetic field, as in
\citet{1997MNRAS.288...63O} and \citet{1998ApJ...499..329O} but with
an isothermal gas.  The boundary conditions are that $B_x$ vanishes at
$z=0$ and has a specified value $B_z\tan i$ at large $z$, together
with the constraint on the surface density.

Eliminating $B_x$, we obtain
\begin{equation}
  \f{\rmd}{\rmd z}\left(\f{1}{2}\Omega^2z^2+\cs^2\ln\rho-\f{2}{3}v_y^2\right)=0
\end{equation}
and
\begin{equation}
  B_z^2\f{\p^2v_y}{\p z^2}+3\Omega^2\mu_0\rho v_y=0.
\end{equation}
For $i<30^\circ$ there exist solutions for which $\rho\to0$ and
$B_x\to B_z\tan i$ as $z\to\infty$, while
$v_y\sim(\tfrac{3}{2}\Omega\tan i)z+v_{y0}$, where $v_{y0}$ is a
constant.  For large $z$ this gives a declining density with
$\ln\rho\sim-(1-3\tan^2i)\Omega^2z^2/2\cs^2$.  For $i>30^\circ$, the
gradient of $v_y$ is sufficiently large that $\rho$ would increase
with $z$ above a certain height, $z=[2\tan
i/(3\tan^2i-1)](-v_{y0}/\Omega)$, and eventually $B_x$ would be
obliged to vary significantly.  This indicates a breakdown of the
static solution and the need for a transcritical outflow.  The
location of the minimum density gives a good indication of the
location of the slow magnetosonic point in the corresponding solution
with an outflow, and the rate of mass loss can be estimated from the
static solution using a correction factor as in
\citet{1998ApJ...499..329O}.

This approximation is expected to be valid when the plasma beta at the
slow point is small, i.e.\ $v_\rma^2\gg\cs^2$.  In this limit the
condition for the slow point reduces to $v_z\approx\cs v_{\rma
  z}/v_\rma$, and the regularity condition~(\ref{regularity}) reduces
to
\begin{equation}
  v_y\approx\f{B_z\Omega z}{2B_x},
\end{equation}
which agrees with what is obtained by assuming an isorotational
configuration and identifying the slow point as the location of the
density minimum of the static approximation as described above.

\subsection{Numerical simulation}
\label{s:dns}

In many ways it is easier to avoid the technical difficulties of
directly computing steady solutions and instead to evolve the system
forwards in time until a steady solution is approached, if it is
stable.  We use the wave speed limiter, usually with $c=10\cs$.  A
branch of steady solutions can be followed quasi-statically by slowing
changing a parameter (such as $B_z$) during the simulation.

To integrate the equations we use tenth-order centred finite
differences in $z$ and a third-order explicit Runge--Kutta method in
$t$.  The method is usually stable when a version of the Courant
condition on the timestep is satisfied.  Boundary conditions are
implemented with the help of ghost zones.  In the case of the free
`outflow' boundary conditions on the velocity, the ghost zones are
filled using linear extrapolation from the computational domain.  If
desired, mass replenishment can be carried out by either the localized
or the distributed method.

\section{Numerical results}

A dimensionless measure of the vertical magnetic field strength is
\begin{equation}
  \f{B_z}{\sqrt{\mu_0\Sigma\cs\Omega}}.
\end{equation}
Without loss of generality, we can choose units such that
$\mu_0=\Sigma=\cs=\Omega=1$.  The remaining parameters are the three
magnetic field components $(B_{x\rms},B_{y\rms},B_z)$ and the three
diffusivities $(\nu,\nu_\rmb,\eta)$, whose values in these units can
be interpreted as Shakura--Sunyaev alpha parameters.

We focus here on jet-launching solutions with $B_{x\rms}=B_z$ and
$B_{y\rms}=0$, for which the inclination of the poloidal magnetic
field lines at the upper boundary is $i=45^\circ$.  The principal
remaining parameter is the vertical magnetic field~$B_z$.  (The main
effect of imposing a non-zero value of $B_{y\rms}$ would be to apply a
magnetic torque to the disc and thereby to drive a mean radial flow.)

\begin{figure}
\centerline{\epsfysize8cm\epsfbox{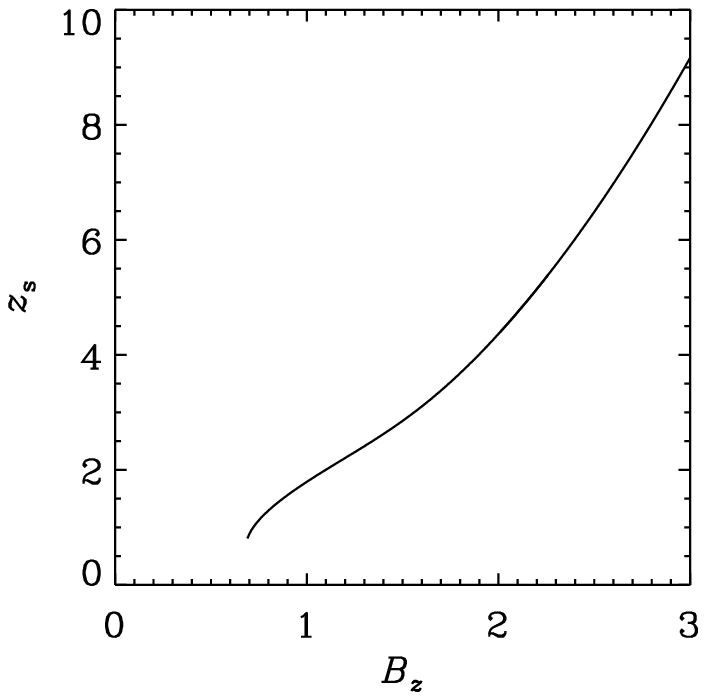}}
\centerline{\epsfysize8cm\epsfbox{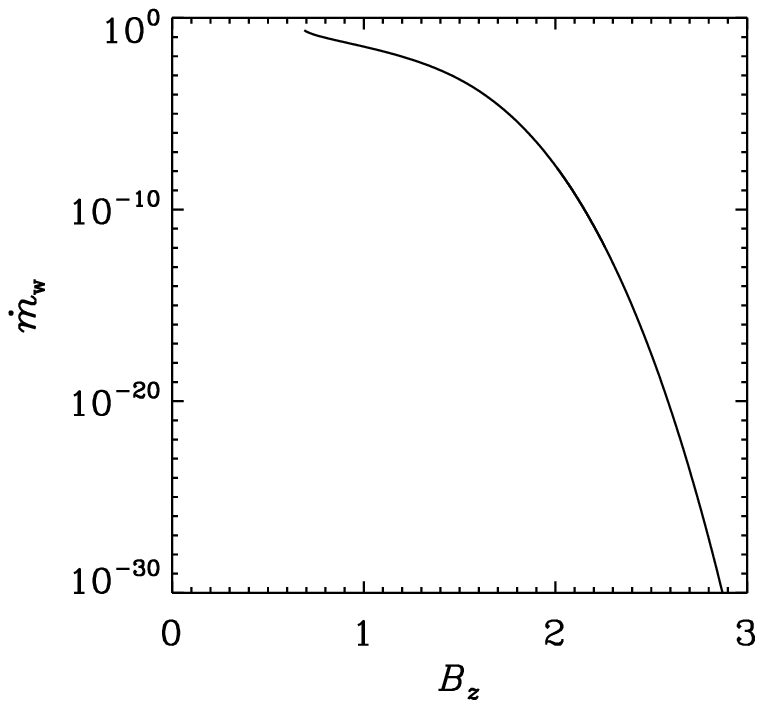}}
\caption{Top: Height of the slow magnetosonic point, in units of
  $\cs/\Omega$, as a function of the vertical magnetic field strength,
  in units of $\sqrt{\mu_0\Sigma\cs\Omega}$, for jet-launching
  solutions with inclination $i=45^\circ$.  Bottom: One-sided
  mass-loss rate per unit area, in units of $\Sigma\Omega$.}
\label{f:results}
\end{figure}

We have computed quasi-steady solutions in ideal MHD as described in
Section~\ref{s:ideal}, using a Runge--Kutta method with adaptive
stepsize.  Mass is replenished at the midplane.  Solutions are
obtained for $B_z\ga0.7$; the height of the slow magnetosonic point
and the rate of mass loss are plotted in Fig.~\ref{f:results}.  In
these solutions the inclination of the poloidal magnetic field is
fixed (i.e.\ $B_x=B_z$) at the slow point.  For $B_z\ga1$ these
quantities are accurately predicted by the simplified version of the
problem described in Section~\ref{s:simplified}.  This behaviour is
consistent with that found by \citet{1997MNRAS.288...63O},
\citet{1998ApJ...499..329O} and \citet{2001ApJ...553..158O}.  A
stronger bending poloidal magnetic field produces a larger radial
Lorentz force, making the disc more sub-Keplerian (in the case of
outward bending) and presenting a larger potential barrier to the
outflow.  The logarithm of the mass-loss rate is proportional to
$-B_z^4$ at large $B_z$.  In practice, the outflow is completely
suppressed when the dimensionless magnetic field strength increases by
a factor of only $2$ or $3$.  However, the maximum value of $\dot
m_\rmw$ plotted in Fig.~\ref{f:results} is about $0.23$, which
corresponds to the disc being emptied on a time-scale of approximately
$0.35$ orbits!

We have also carried out numerical simulations in non-ideal MHD as
described in Section~\ref{s:dns}.  In a typical numerical simulation
we solve the equations in the domain $-5<z<5$ with $1000$ grid points
and with diffusivities $\nu=\nu_\rmb=\eta=0.03$.  A clean start can be
obtained by initializing the simulation with the quasi-steady solution
up to the slow magnetosonic point.  It is straightforward in this way
to obtain strongly magnetized solutions, e.g.\ at $B_z=2$.  In this
regime the poloidal magnetic field does straighten well before $z=5$,
and a steady outflow is achieved, with a rate of mass loss that is
insensitive to the location of the boundary.  With a less careful
initial condition, a permanent epicyclic oscillation is generally
obtained on top of the quasi-steady solution, while other wave modes
damp slowly through viscosity and resistivity.  The epicyclic
oscillation can be avoided only by initializing the horizontal
momentum correctly, but its presence has no effect on the
jet-launching process.

The presence of non-zero diffusivities slightly reduces the rate of
mass loss.  As an example, for $B_z=2$, diffusivities of $0.03$ reduce
the mass loss by about $7\%$ compared to the ideal-MHD solution.

The value of $B_z$ can be decreased slowly and continuously during the
simulation, or the simulation can be restarted periodically with
slightly smaller values of $B_z$.  As this is done, the height of the
slow point decreases and the rate of mass loss increases dramatically.
Eventually the system becomes unstable to the MRI.  If the simulation
maintains perfect symmetry about the midplane, then the first MRI mode
to be observed is expected to be similarly symmetric; if an
antisymmetric seed perturbation is introduced, then an antisymmetric
MRI mode is expected to set in first.  In practice it is difficult to
observe a clean development of the MRI with such an inclined poloidal
field.  The mass loss is so strong in this regime that either the disc
empties very quickly or the details of the replenishment process
become important; depending on how momentum is treated in this
process, a rapid artificial growth of epicyclic motion may occur.
This transitional regime is also problematic because the density of
the outflow is so large that the Alfv\'en point and fast magnetosonic
point approach the disc and the interpretation of the solutions
becomes unclear; there is no longer a force-free region of straight
field lines in which the inclination of the field can be meaningfully
imposed.

Without mass replenishment, the surface density decreases in time,
with the effect that the disc becomes more strongly magnetized: the
value of $B_z/\sqrt{\mu_0\Sigma\cs\Omega}$ increases. If this
parameter starts close to $1$, then rapid mass loss occurs and the
parameter increases quickly until the mass loss abates.  In this way a
burst of ejection occurs.  In a global model of a disc, this condition
might be met at different times at different radii, leading to a
`wave' of ejection.

\section{Conclusion}

The acceleration of an outflow along inclined magnetic field lines
emanating from an accretion disc can be studied in the local
approximation, as employed in the shearing-box model.  By imposing the
inclination of the magnetic field at the vertical boundaries of the
box and resolving the slow magnetosonic point within the computational
domain, appropriate solutions with outflows can be obtained and the
rate of mass loss can be calculated.

This procedure works best when the magnetic field is sufficiently
strong to suppress the magnetorotational instability (MRI).  As found
in previous work, in this regime the mass-loss rate is extremely
sensitive to the dimensionless field strength
$B_z/\sqrt{\mu_0\Sigma\cs\Omega}$, allowing the possibility of an
explosive burst of ejection from any radius in the disc at which the
appropriate field strength is attained.

It would be of interest to determine the stability of these solutions
with respect to perturbations that depend on $x$ and/or $y$, to test
for the instabilities discussed by \citet{1995ApJ...445..337L} and
\citet{1994MNRAS.268.1010L}.  For weaker fields it may be possible to
study the jet launching process in parallel with the MRI.  While that
is a much more demanding problem, the present paper may provide some
useful guidance.

\section*{Acknowledgments}

This research was supported by STFC.  I am grateful to Xuening Bai,
S\'ebastien Fromang, J\'er\^ome Guilet, Geoffroy Lesur and Jim Stone
for useful discussions.

\label{lastpage}


\begin{thebibliography}{}
\bibitem[\protect\citeauthoryear{Balbus 
\& Hawley}{1998}]{1998RvMP...70....1B} Balbus S.~A., Hawley J.~F., 1998, RvMP, 70, 1 
\bibitem[\protect\citeauthoryear{Blandford 
\& Payne}{1982}]{1982MNRAS.199..883B} Blandford R.~D., Payne D.~G., 1982, MNRAS, 199, 883 
\bibitem[\protect\citeauthoryear{Goldreich 
\& Lynden-Bell}{1965}]{1965MNRAS.130..125G} Goldreich P., Lynden-Bell D., 1965, MNRAS, 130, 125 
\bibitem[\protect\citeauthoryear{Hawley, Gammie 
\& Balbus}{1995}]{1995ApJ...440..742H} Hawley J.~F., Gammie C.~F., Balbus S.~A., 1995, ApJ, 440, 742 
\bibitem[\protect\citeauthoryear{Lubow 
\& Spruit}{1995}]{1995ApJ...445..337L} Lubow S.~H., Spruit H.~C., 1995, ApJ, 445, 337 
\bibitem[\protect\citeauthoryear{Lubow, Papaloizou 
\& Pringle}{1994}]{1994MNRAS.268.1010L} Lubow S.~H., Papaloizou J.~C.~B., Pringle J.~E., 1994, MNRAS, 268, 1010 
\bibitem[\protect\citeauthoryear{Miller 
\& Stone}{2000}]{2000ApJ...534..398M} Miller K.~A., Stone J.~M., 2000, ApJ, 534, 398
\bibitem[\protect\citeauthoryear{Ogilvie}{1997}]{1997MNRAS.288...63O} 
Ogilvie G.~I., 1997, MNRAS, 288, 63 
\bibitem[\protect\citeauthoryear{Ogilvie 
\& Livio}{1998}]{1998ApJ...499..329O} Ogilvie G.~I., Livio M., 1998, ApJ, 499, 329 
\bibitem[\protect\citeauthoryear{Ogilvie 
\& Livio}{2001}]{2001ApJ...553..158O} Ogilvie G.~I., Livio M., 2001, ApJ, 553, 158 
\bibitem[\protect\citeauthoryear{Spruit}{1996}]{1996epbs.conf..249S} Spruit 
H.~C., 1996, in Wijers R.~A.~M.~J., Davies M.~B., Tout C.~A., eds, Evolutionary Processes in Binary Stars, Kluwer, Dordrecht, 249
\bibitem[\protect\citeauthoryear{Suzuki 
\& Inutsuka}{2009}]{2009ApJ...691L..49S} Suzuki T.~K., Inutsuka S., 2009, ApJ, 691, L49 
\bibitem[\protect\citeauthoryear{Suzuki, Muto 
\& Inutsuka}{2010}]{2010ApJ...718.1289S} Suzuki T.~K., Muto T., Inutsuka S., 2010, ApJ, 718, 1289 
\end{thebibliography}
\end{document}